\RequirePackage{fix-cm}
\documentclass[twocolumn,epjc3]{svjour3}  

\smartqed 
\RequirePackage{mathptmx}
\RequirePackage{graphicx}
\RequirePackage{rotating}
\RequirePackage{pdflscape}
\usepackage[dvipsnames]{xcolor}
\RequirePackage{latexsym}
\RequirePackage{xspace}
\RequirePackage[numbers,sort&compress]{natbib}
\RequirePackage[colorlinks,citecolor=blue,urlcolor=blue,linkcolor=blue]{hyperref}
\usepackage{amsmath,amsfonts,bm}
\usepackage{wasysym}
\usepackage{upgreek}
\usepackage{subcaption}
\usepackage{comment}
\usepackage{makecell}
\usepackage{hyphenat}
\journalname{Eur. Phys. J. C}
\begin{document}
\newcommand{\DBD}{0$\nu$DBD\xspace}
\newcommand{\onbb}{$0\nu\beta\beta$\xspace}
\newcommand{\nnbb}{$2\nu\beta\beta$\xspace}
\newcommand{\thalf}{$T_{1/2}^{0\nu}$}
\newcommand{\edw}{EDELWEISS\xspace}
\newcommand{\cuore}{CUORE\xspace}
\newcommand{\cupid}{CUPID\xspace}
\newcommand{\cupido}{CUPID-0\xspace}
\newcommand{\cupidmo}{CUPID-Mo\xspace}
\newcommand{\rly}{RLY\xspace}
\newcommand{\ctsper}{counts$/($keV$\cdot$kg$\cdot$yr$)$\xspace}
\newcommand{\Qbb}{$Q_{\beta\beta}$\xspace}
\newcommand{\mbb}{$m_{\beta\beta}$\xspace}

\newcommand{\lmo}{Li$_2$MoO$_4$\xspace}
\newcommand{\enrLMO}{Li$_{2}${}$^{100}$MoO$_4$\xspace}
\newcommand{\ZS}{ZnSe\xspace}
\newcommand{\enrZS}{Zn$^{82}$Se\xspace}
\newcommand{\teo}{TeO$_2$\xspace}
\newcommand{\ntd}{NTD\xspace}

\newcommand\todo[1]{\textcolor{red}{#1}}
\newcommand\edit[1]{\textcolor{blue}{#1}}
\newcommand{\ben}[1]{{\color{Magenta}{[#1] --Ben}}}

\newcommand{\Mo}{$^{100}$Mo\xspace}
\newcommand{\Te}{$^{130}$Te\xspace}
\newcommand{\Se}{$^{82}$Se\xspace}
\newcommand{\TL}{$^{208}\mathrm{Tl}$\xspace}
\newcommand{\Co}{$^{60}\mathrm{Co}$\xspace}
\newcommand{\BEGA}{$\beta/\gamma$\xspace}
\newcommand{\KF}{$^{40}\mathrm{K}$\xspace}
\newcommand{\THO}{$^{232}\mathrm{Th}$\xspace}
\newcommand{\UR}{$^{238}\mathrm{Ur}$\xspace}
\newcommand{\RA}{$^{226}\mathrm{Ra}$\xspace}
\newcommand{\PO}{$^{210}\mathrm{Po}$\xspace}
\newcommand{\PB}{$^{210}\mathrm{Pb}$\xspace}
\newcommand{\FE}{$^{55}\mathrm{Fe}$\xspace}
\newcommand{\MN}{$^{55}\mathrm{Mn}$\xspace}
\newcommand{\BI}{$^{214}\mathrm{Bi}$\xspace}
\newcommand{\PT}{$^{190}\mathrm{Pt}$\xspace}
\newcommand{\K}{$^{40}\mathrm{K}$\xspace}
\newcommand{\al}{$\alpha$\xspace}
\newcommand{\be}{$\beta$\xspace}
\newcommand{\ga}{$\gamma$\xspace}
\newcommand{\xr}{X-ray\xspace}
\newcommand{\ky}{kg$\times$yr\xspace}
\newcommand{\ckky}{counts/(keV$\times$kg$\times$yr)\xspace}
\newcommand{\cpdkk}{\un{cpd/keV/kg}\xspace}
\providecommand*{\un}[1]{\ensuremath{\mathrm{~#1}}}
\hyphenation{mono-ener-getic}

\title{Optimized filtering for pulse-shape based pile-up rejection applied to $0\nu\beta\beta$ search with $^{100}$Mo}

\author{
V.~Berest\thanksref{CEA-IRFU,f0}\and
M.~Buchynska\thanksref{IJCLab}\and
P.~Carniti\thanksref{INFN-Milano, Milano} \and
A.~Giuliani\thanksref{IJCLab}\and
C.~Gotti\thanksref{INFN-Milano, Milano}\and
H.~Khalife\thanksref{CEA-IRFU,f1}\and
P.~Loaiza\thanksref{IJCLab}\and
P.~de~Marcillac\thanksref{IJCLab}\and
C.~Nones\thanksref{CEA-IRFU}\and
M.~Pageot\thanksref{CEA-IRFU,e1}\and
E.~Olivieri\thanksref{IJCLab}\and
G.~Pessina\thanksref{INFN-Milano, Milano}\and
D.~V.~Poda\thanksref{IJCLab}\and
J.~A.~Scarpaci\thanksref{IJCLab}\and
B.~Schmidt\thanksref{CEA-IRFU}\and
A.~S.~Zolotarova\thanksref{CEA-IRFU}
}

\thankstext{f0}{Now at University of California, Berkeley, CA, USA}
\thankstext{f1}{Now at IJCLab, Orsay, France}
\thankstext{e1}{e-mail: mathieu.pageot@cea.fr}
\institute{IRFU, CEA, Universit\'{e} Paris-Saclay, F-91191 Gif-sur-Yvette, France  \label{CEA-IRFU} \and 
Universit\'{e} Paris-Saclay, CNRS/IN2P3, IJCLab, 91405 Orsay, France \label{IJCLab} \and
INFN, Sezione di Milano-Bicocca, I-20126 Milano, Italy \label{INFN-Milano}  \and 
Dipartimento di Fisica, Universit\`{a} di Milano-Bicocca, I-20126 Milano, Italy \label{Milano}
}

\date{Received: date / Accepted: date}

\maketitle
\begin{abstract}
Pile-up events, arising from the partial or complete temporal overlap of distinct signals, represent a major challenge in many areas of experimental physics where rare or low-rate processes are targeted. If not properly identified, pile-up can distort reconstructed observables, degrade energy resolution, and generate backgrounds that mimic genuine events of interest. This work presents an algorithm to obtain an optimized digital filter for the discrimination of pile-up events for detectors with known signal response and stationary noise power spectral density.
It is developed in the context of the search for neutrinoless double beta decay with cryogenic \enrLMO detectors like CUPID, where pile-up induced background from $^{100}$Mo \nnbb is expected to be the leading background contribution.  For this application, the new filter discriminant reduces the pile-up induced background (at 90\% efficiency) by 31\%, compared to an analysis with a reference method previously presented in Eur. Phys. J. C 83(5), 373 (2023).
While the discussion is grounded in cryogenic calorimetric detectors, the concepts and methods described are broadly applicable to a wide class of detector technologies and experimental contexts.
\end{abstract}

\section{Introduction}

Modern experimental setups frequently operate in regimes where the event rate approaches or exceeds the characteristic response time of the detector. Under these conditions, multiple interactions can contribute to a single recorded waveform, complicating the reconstruction of individual events \cite{Pileup_Pulse_Height_Spectra,Pileup_photon_counting_xray_detectors,Pileup_HOLMES,Pileup_liquid_scintillator,Pileup_Plasma}. Such signal overlap, called pile-up, can lead to distorted reconstructed observables, degraded energy resolution, efficiency loss, dead time, and the generation of background events that mimic genuine signals of interest \cite{Knoll,HOLMES_optimum_filter,CUPID_pileup_NTL}.

This issue is particularly critical for slow detectors such as cryogenic macro-calorimeters, characterized by intrinsic thermal response times spanning from a few milliseconds to several hundreds of milliseconds. Experiments based on this technology, including CUPID \cite{CUPID}, AMoRE \cite{AMoRE_II}, ECHO \cite{ECHO}  and HOLMES \cite{HOLMES}, are therefore especially sensitive to pile-up effects. Among them, CUPID (CUORE Upgrade with Particle IDentification) faces particularly stringent requirements, as it combines slow thermal signals with the need for extremely low background levels in a rare-event search.
CUPID aims to search for neutrinoless double beta decay (\onbb), a hypothetical nuclear process whose observation would establish the Majorana nature of neutrinos and demonstrate the violation of lepton number conservation \cite{Lepton_Number_Violation}. The decay can proceed through several beyond-the-Standard-Model mechanisms \cite{BSM}, and its discovery would provide access to several new physics models at mass scales beyond the reach of current accelerators. In its simplest and most widely discussed model, the decay is mediated by the exchange of light Majorana neutrinos \cite{Majorana, Furry}. In this framework the decay rate is directly related to the effective Majorana neutrino mass (the linear combinations of neutrino mass eigenvalues), providing a complementary means to assess the absolute neutrino mass scale in addition to  cosmological inference \cite{Planck} and kinetic mass measurements in beta decay \cite{KATRIN}. Moreover, \onbb\ may be connected to the matter–antimatter asymmetry in models where baryogenesis \cite{Barygenesis} proceeds through leptogenesis in the Early Universe \cite{Matter_creation}.

Experimentally, \onbb\ is a nuclear transition $(Z,N) \rightarrow (Z+2, N-2) + 2e^{-}$, which transforms a nucleus into its isobar with the emission of two electrons. It may be observable in nuclei where single beta decay is forbidden (or strongly suppressed) as it has a very clean signature of a mono-energetic peak at the transition energy $Q_{\beta\beta}$. 
However, it has not been observed yet with existing experiments probing half-lives of $10^{25}$ to $10^{26}$\,yrs \cite{KamLAND, LEGEND_200, CUORE_last_result}.  Achieving the required sensitivities requires detectors with excellent energy resolution and extremely low background levels.  Cryogenic calorimeters are among the most sensitive technologies for this search, providing leading constraints on several candidate isotopes: \Te \cite{CUORE_last_result}, \Mo \cite{AMoRE_1, CUPID_Mo}, and \Se \cite{CUPID_0}. Among these, \Mo-based cryogenic calorimeters have received particular interest in the community due to their excellent detector performance along with scientifically and economically favorable characteristics. \Mo has a high \Qbb-value which results in a large phase-space factor \cite{Phase_space_factors} and which reduces radiogenic backgrounds as the expected peak energy lies above the most intense $\gamma$-ray lines from natural radioactivity. Existing technologies for enrichment further allow for a reasonable scaling for large detector arrays. In addition, the scintillation of \lmo crystals enables simultaneous heat and light detection when coupled to dedicated light detectors (LDs). The combination of these two channels provides particle identification, allowing efficient rejection of $\alpha$-induced background events. For these reasons, \Mo embedded in scintillating \lmo crystals has been selected as the isotope of choice for next-generation searches in CUPID \cite{CUPID} and AMoRE-II \cite{AMoRE_II}. 

However, \Mo's favorable kinematics comes with one disadvantage from the Standard Model allowed two-neutrino double beta decay (\nnbb), which dominates the detector rate and can lead to unidentified pile-up constituting an irreducible source of background. With a measured half-life of $7.07 \times 10^{18}$~yr for $^{100}$Mo \cite{CUPID_Mo_2nbb}, the temporal overlap of two \nnbb\ events can produce pile-up signals that mimic the \onbb\ at 3034 keV \cite{DoubleBeta_Spectrum}. Based on current detector performance \cite{CUPID} and analysis methods \cite{CUPID_pileup,CUPID_pileup_NTL}, CUPID expects this contribution to dominate the overall background budget, with a design target of 50\% of the total background arising from pile-up and a target pile-up background index of $5 \times 10^{-5}$~counts/keV/kg/year \cite{CUPID}.
Significant experimental efforts within and beyond CUPID are being devoted to the mitigation of pile-up at the detector level. These include the development of faster thermal sensors such as Transition Edge Sensors (TESs) in CUPID \cite{CUPID_TES,CUPID_TES_LD} and Metallic Magnetic Calorimeters (MMCs) in AMoRE \cite{AMoRE_MMC} or the further improvement of CUPID's detector technology. In CUPID the scintillating cryogenic calorimeters are coupled to LDs operated with Neganov–Trofimov–Luke (NTL) amplification \cite{Neganov_Trofimov,Luke}, which provides improved timing and pulse-shape information from the LD \cite{NTL_detector} with rise times of $\sim$~[0.5, 0.8]~ms, substantially faster than the $\sim 10$~ms rise times of the \lmo calorimeter signals. Such advances aim at reducing the effective pulse duration, improving  the signal-to-noise ratio (SNR), and enhancing the ability to resolve closely spaced interactions. 
Nevertheless, in large-scale cryogenic experiments targeting isotopes with high \nnbb rates, residual pile-up remains unavoidable and ultimately sets stringent requirements on the analysis techniques.
As a consequence, the identification and rejection of pile-up events through pulse-shape discrimination remains a central challenge and has been the subject of several previous studies \cite{Pileup_HOLMES,CUPID_pileup_old,CUPID_pileup_ML,ZnMoO_pileup}. It constitutes a complementary handle on the software and analysis side to further improve experimental sensitivity, and motivates the development of a , quantitative, and general framework for the modeling, identification, and rejection of pile-up events in cryogenic calorimetric detectors, which is the focus of the present work.

\section{Analytical framework}
\label{sec:framework}

\begin{figure}[!ht]
    \centering
    \includegraphics[width=0.8\linewidth]{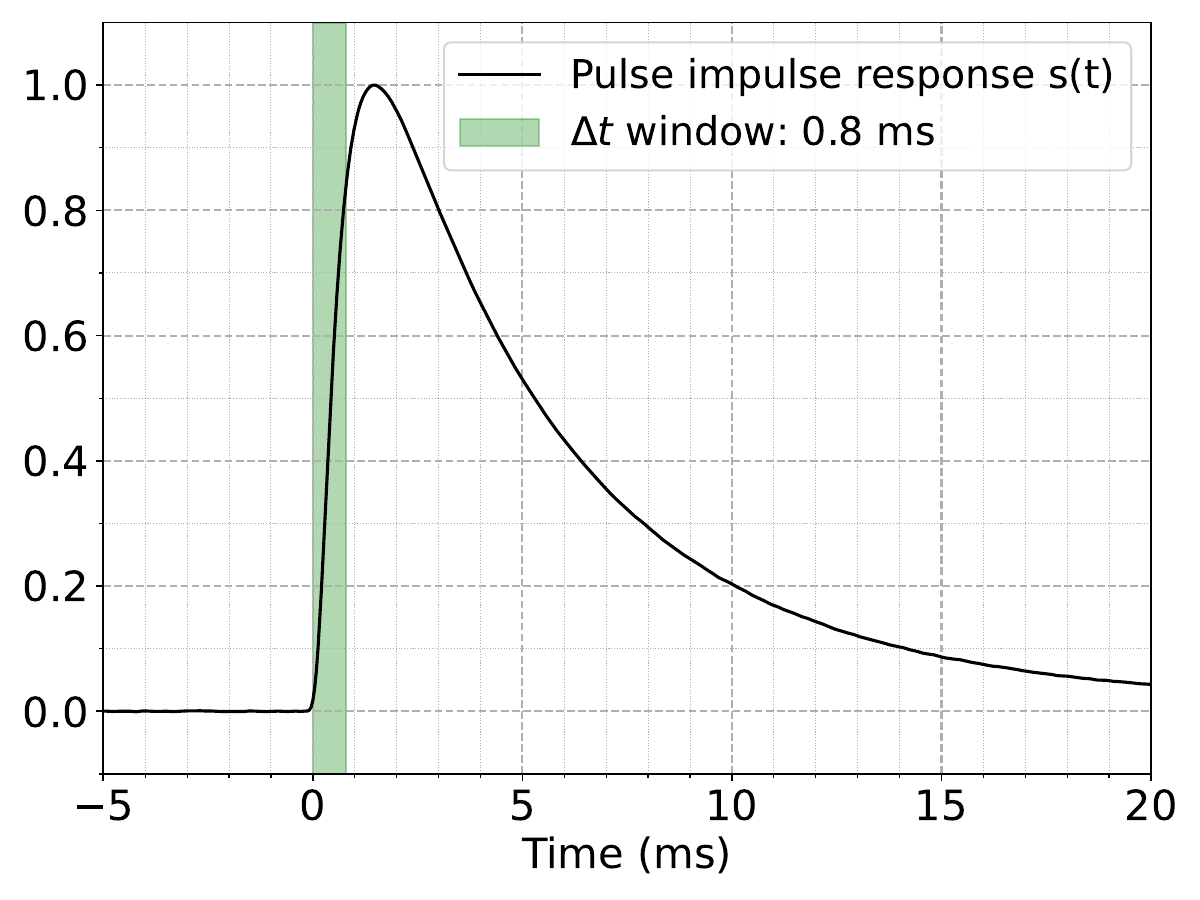}
    \caption{Example of the pulse impulse response $s(t)$ of a LD used in this study, together with the range of pile-up time separations investigated.}
    \label{fig:meanpulse}
\end{figure}

In the following we assume a fully known, energy independent impulse response of the detector to an energy deposition.
The measured signal can be expressed as the sum of this single pulse impulse response $s(t)$ , illustrated in Fig.~\ref{fig:meanpulse}, and stochastic noise processes $n(t)$,
\begin{equation}
    x(t) = A\,s(t) + n(t),
\end{equation}
where $A$ denotes the pulse amplitude proportional to the energy deposition in the detector.

\subsection{Pile-up model}
A pile-up event arises from the temporal overlap of two pulses, which we parametrize as:
\begin{equation}
    x_{\mathrm{PU}}(t) = (1-r) A s(t) + r A s(t+\Delta t) + n(t),
\end{equation}
where $r$ denotes the relative amplitude and $\Delta t$ the time separation between the pulses. 

\begin{figure}[ht]
\centering
\begin{subfigure}[t]{0.45\textwidth}
  \centering
  \includegraphics[width=.99\linewidth]{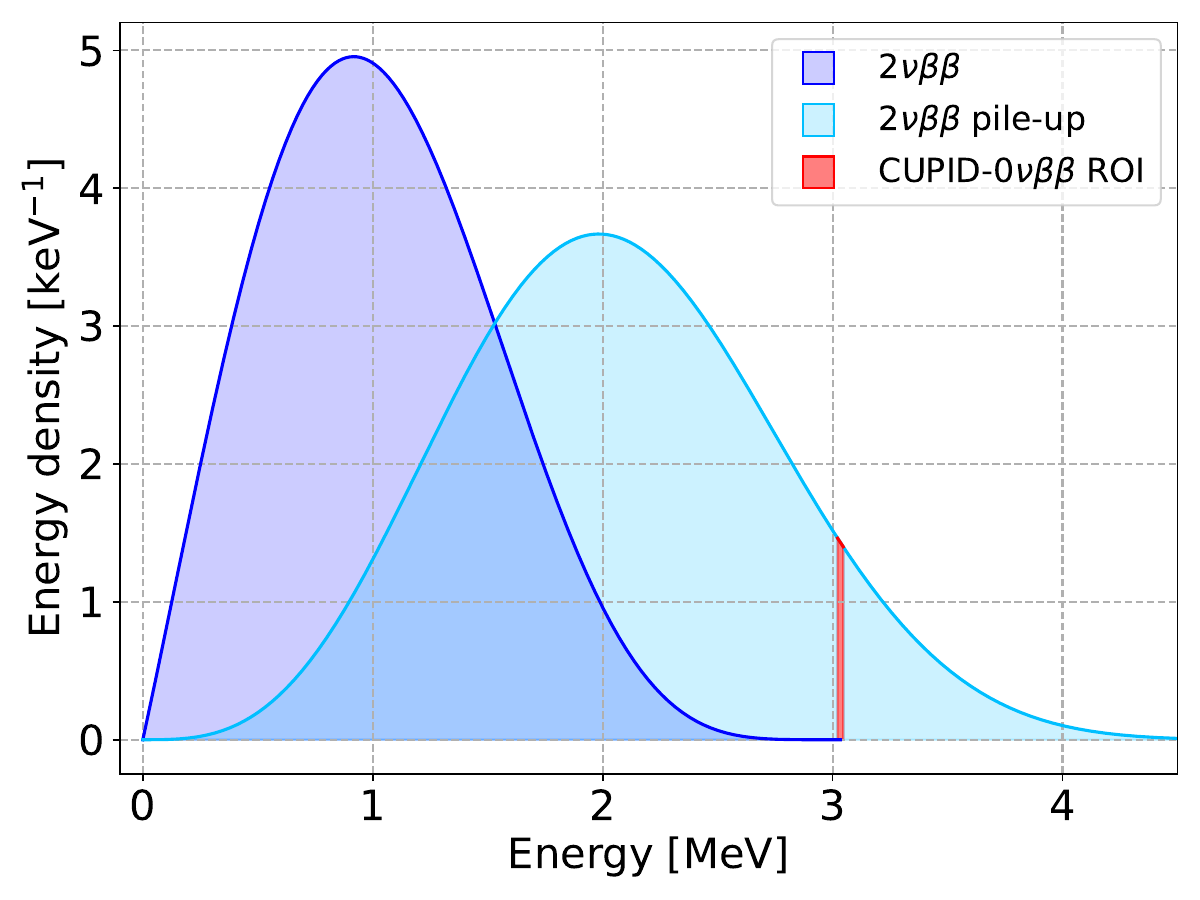}
\end{subfigure}
\hfill
\begin{subfigure}[t]{0.45\textwidth}
  \centering
  \includegraphics[width=.99\linewidth]{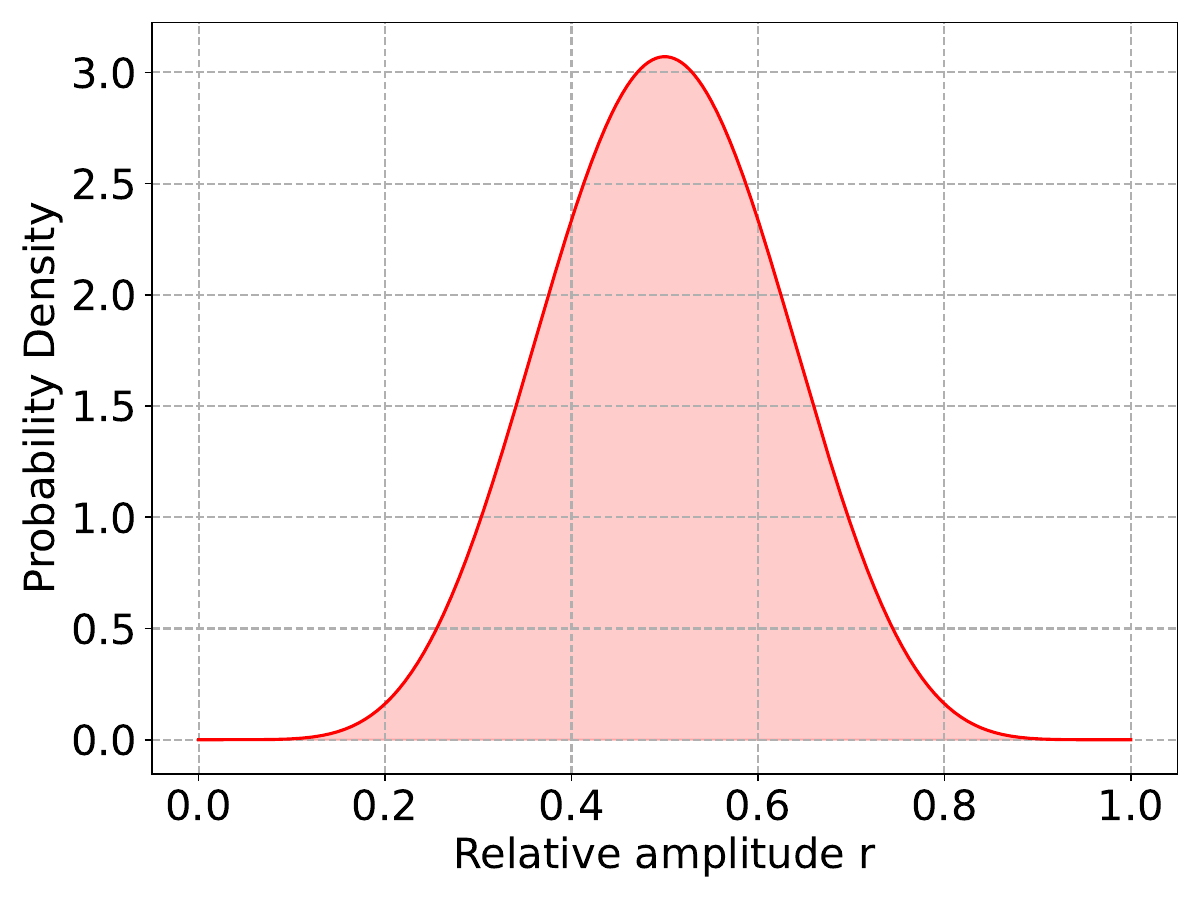}
\end{subfigure}
\caption{(top) Approximation of the $2\nu\beta\beta$ energy spectrum of $^{100}$Mo \cite{DoubleBeta_Spectrum}: 
$E \cdot (3.034-E)^5 \cdot (E^4 + 10 E^3 + 40 E^2 + 60 E + 30)$, where $E$ is the energy in keV together with the corresponding $2\nu\beta\beta$ pile-up energy spectrum. The CUPID-$0\nu\beta\beta$ ROI is highlighted.
(bottom) Distribution of the relative amplitude $f_R(r)$ for event pairs with total energy $Q_{\beta\beta} = 3.034~\mathrm{MeV}$. 
This figure illustrates the \nnbb\ decay spectrum and the resulting distribution of relative amplitudes $r$ for pile-up events.}
\label{fig:r_pdf}
\end{figure}

In the example case of a \onbb search with \Mo their distributions are approximated by the \nnbb decay spectrum and its detector event rate, respectively.  Figure~\ref{fig:r_pdf} shows an approximation of the $2\nu\beta\beta$ energy spectrum of $^{100}$Mo \cite{DoubleBeta_Spectrum} and the corresponding distribution of the relative amplitude, $f_R(r)$, with sums of energy equal to 3034 keV.  For this study it is assumed that only $2\nu\beta\beta$ contributes to the pile-up background given the predominance of those events in the low radioactivity environment of future or past experiments \cite{CUPID_pileup_NTL,CUPID_Mo, ZnMoO_pileup}. This study takes advantage of the fast response of the LD, whose energy is scaled by the amount of light registered on the LD for an energy deposit in the \lmo{}. Historically, this is referred to as light yield for cryogenic \lmo{} bolometers, but it includes the detection efficiency with a typical value of 0.36 keV/MeV in an open structure like CUPID \cite{CUPID_GBPT}. However, this scaling factor does not affect the distribution of the relative amplitude, and therefore does not modify ($f_R(r)$).  Within this framework, we restrict the discussion to \emph{unresolved} pile-up events, defined as pulses separated by time intervals shorter than the signal rise time (taken as the 10\%--90\% rise interval of the pulse), i.e.\ $\Delta t < 0.8$~ms in the LDs, for which standard filtering techniques combined with $\chi^2$-based pile-up rejection lose discrimination power.
Within this time range, the pulse separation is assumed to be uniformly distributed, $f_{\Delta t}(\Delta t) = \frac{1}{T_\mathrm{max}}$ for $\Delta t \in [0, T_\mathrm{max}]$ and $T_\mathrm{max}=0.8$~ms.
This assumption is justified by the fact that the time window of interest for unresolved pile-up (0.8~ms) is much shorter than the mean inter-event arrival time, $\lambda = 340$~s, corresponding to a $2\nu\beta\beta$ decay rate of 2.94~mHz in a 280~g \lmo crystal with 95\% $^{100}$Mo enrichment \cite{CUPID}, using the half-life measured by CUPID-Mo \cite{CUPID_Mo_2nbb}.

\subsection{Discriminant parameter}

In order to identify and reject pile-up events, we adopt a frequency-domain description of the signal~\cite{Fourier}. The noise $n(t)$ is assumed to be a stationary, zero-mean Gaussian stochastic process, fully characterized by its power spectral density $\mathcal{P}_N(\omega)$.

We denote by $\mathcal{F}$ the Fourier transform and by $\mathcal{F}^{-1}$ its inverse. Using the notation
\begin{equation}
    \hat{s}(\omega) = \mathcal{F}\{s(t)\} = \int_{-\infty}^{+\infty} s(t)\, e^{-i\omega t}\, dt,
\end{equation}
we also define $\hat{n}(\omega)$ as the Fourier transform of the noise $n(t)$:
\begin{equation}
    \hat{n}(\omega) = \mathcal{F}\{n(t)\} = \int_{-\infty}^{+\infty} n(t)\, e^{-i\omega t}\, dt.
\end{equation}
The power spectral density is then defined as the average of the squared Fourier amplitudes over multiple noise traces:
\begin{equation}
    \mathcal{P}_N(\omega) = \Big\langle \big|\hat{n}(\omega)\big|^2 \Big\rangle.
\end{equation}

In the frequency domain, the measured signal can then be written as
\begin{equation}
    \hat{s}_{\mathrm{measured}}(\omega) = A\,\hat{s}(\omega) + \hat{n}(\omega).
\end{equation}

To construct the pile-up filter and discriminant, we start from the classical optimum filtering approach~\cite{Gatti_Manfredi}, with the transfer function
\begin{equation}
    H(\omega) = \frac{1}{\mathcal{N}} \frac{\hat{s}^{*}(\omega)}{\mathcal{P}_N(\omega)},
\end{equation}
which maximizes the SNR. The normalization constant $\mathcal{N}$ is chosen such that
\begin{equation}
    \int H(\omega)\,\hat{s}(\omega)\, d\omega = 1 .
\end{equation}

To separate pile-up from single events, we define an amplitude ratio-based parameter as previously used in \cite{CUPID_pileup}, but with a modified definition of the two amplitude estimators:

\begin{equation}
    Y[m_1,m_2](\Delta t,r) = \frac{X[m_1](\Delta t,r)}{X[m_2](\Delta t,r)},
\end{equation}
where
\begin{equation}
    X[m_i](\Delta t,r) = \max \Big( \operatorname{Re} \big[ \mathcal{F}^{-1}( m_i(\omega) H(\omega) \hat{x}(\omega,\Delta t,r) ) \big] \Big),
\end{equation}
and the $m_i(\omega)$ are positive real weighting functions to be optimized normalized such that $\mathcal{F}^{-1}( m_i(\omega) H(\omega) \hat{s}(\omega))=1$ (i.e. not to change the normalization of the optimum filter). 
Similar to the standard optimum filter amplitude,the $X[m_i]$ are also expected to closely follow a Gaussian distribution when evaluating a mono-ener\hyp{}getic pulse response with real noise data.
In this context, $X[m_1]$ serves as a pile-up-sensitive energy estimator, while $X[m_2]$ provides a more stable energy estimator under pile-up used for normalization. In the following, we derive the loss function how to optimize $m_i$, which is later exemplified for the case of pile-up from \nnbb in Sec.~\ref{sec:results}.

The residual \emph{unidentified pile-up rate} $\mathcal{M}(\Delta t,r)$ is defined as the probability that a pile-up event is incorrectly identified as a single pulse:
\begin{equation}
    \mathcal{M}[m_1,m_2](\Delta t,r) = P\big(Y[m_1,m_2](\Delta t,r) > Y_\text{cut}[m_1,m_2]\big),
\end{equation}
where $Y_\text{cut}$ is the threshold chosen to achieve a desired single-pulse acceptance $\epsilon$.

Assuming that $Y[m_1,m_2](\Delta t,r)$ is approximately Gaussian  \cite{Normal_ratio} (see following discussion),
\begin{equation}
    Y(\Delta t,r) \sim \mathcal{N}(\mu_Y(\Delta t,r),\sigma_Y(\Delta t,r)),
\end{equation}
the misidentified pile-up rate can be written explicitly as
\begin{equation}
    \mathcal{M}[m_1,m_2](\Delta t,r) = 1 - \Phi\left( \frac{Y_\text{cut}[m_1,m_2] - \mu_Y[m_1,m_2](\Delta t,r)}{\sigma_Y[m_1,m_2](\Delta t,r)} \right),
\end{equation}
with $\Phi$ the standard normal cumulative distribution function, and
\begin{equation}
    Y_\text{cut}[m_1,m_2] = 1 - N_\sigma \sigma_{Y,0}[m_1,m_2], \quad N_\sigma = \Phi^{-1}(\epsilon),
\end{equation}
where $\sigma_{Y,0}$ is the standard deviation of $Y$ for single pulses.

Complete expressions for $\mu_Y(\Delta t,r)$, $\sigma_Y(\Delta t,r)$ are derived in~\ref{app:stat_moment}.

The \emph{global pile-up misidentification rate} is obtained by integrating over the distributions of amplitude ratio $r$ and time separation $\Delta t$:
\begin{equation}
    \langle \mathcal{M} \rangle[m_1,m_2] = \iint \mathcal{M}[m_1,m_2](\Delta t,r)\, f_R(r)f_{\Delta t}(\Delta t)\, dr\, d\Delta t.
\end{equation}
This global misidentification rate $\langle \mathcal{M} \rangle$ serves as the cost function to be minimized with respect to $m_1$ and $m_2$ to provide an optimal amplitude ratio-based filter for pile-up discrimination.

The choice of an amplitude ratio as discriminator is not guaranteed to be optimal. However, it has several advantages; particularly, it reduces the sensitivity to variation in amplitude. It also allows to exploit a strong correlation between $X[m_1]$ and $X[m_2]$,  thereby reducing the variance \( \sigma_Y \) and improving discrimination as highlighted in the more explicit derivations in the supplemental material (\ref{app:stat_moment}). Similar benefits have been reported in \cite{CUPID_pileup}, where a ratio-based Pulse Shape Discriminator (PSD) was shown to outperform other methods, including $\chi^2$-based discriminators.

The Gaussian approximation for $Y(\delta t, r)$ is justifiable for the use in \onbb search, where present detectors provide Gaussian distributed variables \( X[m_i] \) with standard deviations and covariance small compared to their non-zero mean values.

The method assumes stationary pulse shapes and noise as generally assumed in the data processing of cryogenic detectors for rare event search \cite{CUORE_analysis, CUPID_0_Analysis, CUPID_Mo_analysis}. It provides a set of optimal linear filters for a ratio-based PSD, when the nominal amplitude is used and it is robust against small amplitude deviations.

\

\section{Implementation}
\label{sec:implementation}

The pile-up discrimination framework described in Sec.~\ref{sec:framework} has been implemented in Python using \texttt{PyTorch},  making use of the automatic differentiation to optimise the weighting functions via gradient descent \cite{Pytorch}.

The reference pulse shape $\hat{s}(\omega)$ is obtained by averaging a large ensemble of pulses within a selected energy region after aligning them with an optimum filter. The noise power spectrum $\mathcal{P}_N(\omega)$ is derived from randomly triggered data windows, rejecting pulse contamination according to their root mean square (RMS). For each window, the Fourier transform is computed, and $\mathcal{P}_N(\omega)$ is obtained by averaging the squared magnitudes of these spectra.
The frequency-domain weighting functions $m_1(\omega)$ and $m_2(\omega)$ are treated as \emph{trainable parameters} vectors in \texttt{PyTorch}. The number of trainable parameters is directly determined by the length of the processing window, as the weighting functions are discretized over the corresponding frequency bins. These parameters are optimized to minimize the pile-up misidentification rate  given a fixed acceptance for single pulses.

A discrete grid of time offsets $\Delta t$ and amplitude ratios $r$ is used to represent possible pile-up events.
Since exchanging the two pulses is equivalent to the transformation $r \rightarrow 1-r$, the pile-up model is symmetric about $r=0.5$. The computation can therefore be restricted to the interval $r\in[0,0.5]$, with the corresponding normalization adjusted accordingly, reducing the computational cost without loss of information.
For each $(\Delta t,r)$ pair, the misidentification rate  $\mathcal{M}[m_1,m_2](\Delta t,r)$ is computed using the Gaussian approximation of the discriminant parameter (Sec.~\ref{sec:framework}). To reach subsample resolution, a 1D cubic Hermite interpolation \cite{Interpol_Hermite} is applied on the filtered signals before taking their maximum. The global cost function is then obtained by summing over the grid (given a proper normalization of the distributions):
\begin{equation}
    \langle \mathcal{M} \rangle[m_1,m_2] = \sum_{\Delta t,r} \mathcal{M}[m_1,m_2](\Delta t,r)\, f_R(r)\, f_{\Delta t}(\Delta t)\,dr\, d\Delta t,
\end{equation}
which represents the expected probability of misclassifying pile-up events as single pulses.

The weighting functions $m_1$ and $m_2$ are updated iteratively using \texttt{PyTorch}’s automatic differentiation. 
At each step, gradients $\partial \langle \mathcal{M} \rangle/\partial m_i$ are computed and the parameters are adjusted according to a chosen optimizer (e.g., Adam or SGD) until convergence. 
The result is an \emph{optimized frequency-domain filter} that minimizes pile-up misidentification rate --- maximizes pile-up rejection --- for a fixed single pulse efficiency.

The pipeline is modular and fully compatible with GPU acceleration, enabling efficient evaluation over large $(\Delta t,r)$ grids.

\section{Performance assessment}
\label{sec:performance}

The proposed pile-up discrimination framework was tested through performance evaluation by using data from the CROSS experiment (Cryogenic Rare-event Observatory with Surface Sensitivity)~\cite{CROSS}.
In particular, we used data from a measurement campaign, where the CROSS collaboration operated a tower of 6 scintillating \lmo crystals and 4 \teo crystals coupled to Ge LDs operated with NTL amplification, providing a suitable test bench for CUPID-like conditions \cite{CROSS_NTL}. The purpose of this analysis is to exploit this data set to evaluate the performance of the proposed method and assess an optimal operation setting for the CUPID experiment.

\subsection{Detector configuration}
The full details about the detector configuration are given in~\cite{CROSS_NTL} and we will only provide a summary here. The detector array was comprised of 10 crystals coupled to NTL-assisted LDs.  
The study of pile-up rejection is performed on the LDs only, as their faster response dominates the pile-up rejection capability. Eight out of the ten available LDs were selected rejecting detectors 1 and 7 due to a broken electrical connection and a misbehaving detector performance \cite{CROSS_NTL}. 
The data were collected at 22~mK with an 80~V electrode bias for NTL amplification. 
The acquisition was set with a 10~kHz digitization rate and a 6th-order Bessel filter with a cutoff frequency of 2.5~kHz. The measured pulse rise times (10--90\%) range from 0.5 to 0.8~ms. To estimate the expected SNR for CUPID, the measured CROSS data were rescaled to account for differences in light yield (the measured scintillation light per unit of deposited energy in the crystal), going from 0.3~keV/MeV to 0.36~keV/MeV and electrode coverage (from 56\% to 75\%) between the two experiments. This scaling discussed as configuration III (scaling of x1.6, reaching a light yield of 0.36 keV/MeV) in \cite{CROSS_NTL} provides a SNR range of 79 to 239 at the Q-value.
This variation in detector working conditions and response allows assessing the robustness of the proposed discrimination method across different working regimes.

\subsection{Evaluation dataset generation}
The test data simulates pile-up events by combining simulated signals with real noise collected from experimental detectors \cite{CROSS_NTL} as developed previously in \cite{CUPID_pileup}. 
High-quality single-pulse templates extracted from calibration data are scaled and added to noise  to generate both single and pile-up events.  
Pile-up events are produced by a random sampling of the continuous amplitude ratio distribution for pile-up in the \lmo detector that produces a sum energy within the range $E$ = [3019, 3049] keV.
The amplitudes are translated into respective LD signals including a statistical Poisson fluctuation of the number of photons and the LD's sensitivity scaled to the CUPID detector parameters. 
For the present study we assume that the noise power spectrum for these detectors can be approximated by the ones measured in the CROSS cryostat. This is rooted in the fact that the CROSS setup serves to validate a close to final electronics/instrumentation chain of CUPID. However, a major cryostat upgrade is expected for CUPID, which will require a dedicated study of its vibrational noise environment and which has to be considered when interpreting the quoted results versus the CUPID background targets. 
We note that a potential impact of a mis-reconstructed energy in the \lmo calorimeter at the level of 5 keV, the design resolution (FWHM) of CUPID translates into a negligible shift in the mean number of photons in comparison and is neglected. 
 Finally, the two LD signals are superimposed to a random noise trace, which contributes to the effect of the finite detector baseline resolution. A random sub-sample shift of the initial pulse-onset and a uniform random inter-arrival time within the range $[0, 0.8]$~ms complete the generation of the evaluation data set.  

\subsection{Evaluation procedure}
For each detector, the analytical model is parametrized using the measured average pulse $\hat{s}(\omega)$ and the noise power spectral density $\mathcal{P}_N(\omega)$. To avoid introducing artificial correlations between the injected signals and the filter training, the pulse template used for optimization is recomputed from the injected dataset rather than reusing the injection template.
The optimization process from Sec.~\ref{sec:implementation} is run separately for each detector to produce individual weighting functions $m_1(\omega)$ and $m_2(\omega)$ for every detector. 
The resulting discriminant parameter $Y[m_1,m_2]$ is then evaluated for both injected single-pulse and pile-up events to assess separation performance, expressed in terms of rejection power $\mathcal{R} = N_{\mathrm{rej}} / N_{\mathrm{inj}}$. Here, $N_{\mathrm{inj}}=16000$ is the total number of injected pile-up pulses and $N_{\mathrm{rej}}$ denotes the number of pile-up events whose discriminant value $Y$ falls below the threshold defined by the $n$-th percentile of the single-pulse distribution, where $n$ is fixed by the target single-pulse efficiency. The BI in the ROI, usually stated in  cts/keV/kg/yr,  quantifying the rate of spurious events that can mimic the \onbb signal is derived using the following formula \cite{CROSS_NTL}:
\begin{equation}
    \text{BI} = \rho_E\cdot\frac{\Gamma^2}{m} \cdot(1-\mathcal{R}) \cdot \Delta t_{\mathcal{R}=100\%} \cdot T_{yr},
\end{equation}
where $\Gamma = 2.94 \times 10^{-3}~\mathrm{s^{-1}}$ is the \Mo $2\nu\beta\beta$ event rate in the \Mo-enriched \lmo crystal, 
$m = 0.28~\mathrm{kg}$ is the crystal mass, 
$\rho_E = 3.41 \times 10^{-4}~\mathrm{keV^{-1}}$ \cite{CUPID_pileup} is the probability density of observing the summed energy of two \Mo $2\nu\beta\beta$ events at $Q_{\beta\beta} = 3034~\mathrm{keV}$ (neglecting energy mis-reconstruction), 
$\Delta t_{\mathcal{R}=100\%} = 0.8~\mathrm{ms}$ is the estimated pulse separation guaranteeing complete pile-up identification, 
$\mathcal{R}$ is the global rejection power for pile-up events below $\Delta t_{\mathcal{R}=100\%}$, and 
$T_{yr}= 3.154 \times 10^7~\mathrm{s/yr}$ is the number of seconds in a year. In this equation, the term $\Gamma^2 \Delta t_{\mathcal{R}=100\%} = 6.9\times10^{-9}~\mathrm{Hz}$ corresponds to the maximum rate of $2\nu\beta\beta$ pile-up events in a CUPID crystal that could remain unresolved.

\subsection{Results}
\label{sec:results}
\begin{figure}[ht]
    \centering
    \includegraphics[width=0.8\linewidth]{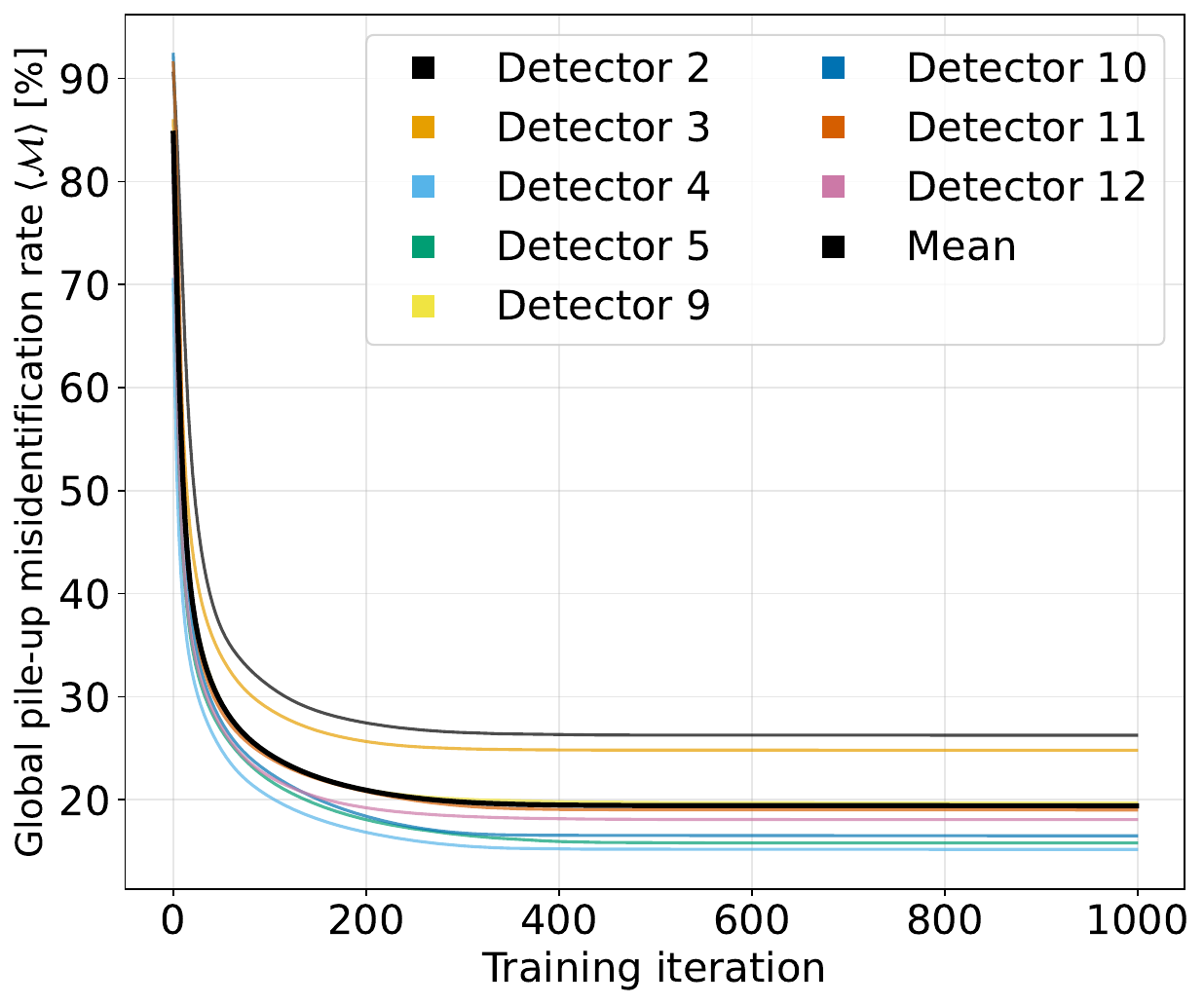}
    \caption{Training evolution of the cost function $\langle \mathcal{M} \rangle[m_1, m_2]$.}
    \label{fig:training_J}
\end{figure}

Figure~\ref{fig:training_J} shows the evolution of the cost function $\langle \mathcal{M} \rangle[m_1, m_2]$ during iterative gradient descent for the different detectors. The optimization process shows stable convergence across all detectors, with the cost function decreasing rapidly in the initial iterations before reaching a plateau, confirming the robustness of the gradient-descent approach in detector-specific conditions.
The training time required for a single detector is less than 17~s, and the usage of GPU memory remains below 1~GB per channel. In a large-scale computing facility such as the National Energy Research Scientific Computing Center (NERSC) \cite{NERSC}, which provides GPU- accelerated resources, a fully parallel training of the 1653 CUPID detectors would require only about 0.5\% of the total available GPU resources (328~TB of memory).

\begin{figure}[!h]
    \centering
    \includegraphics[width=0.8\linewidth]{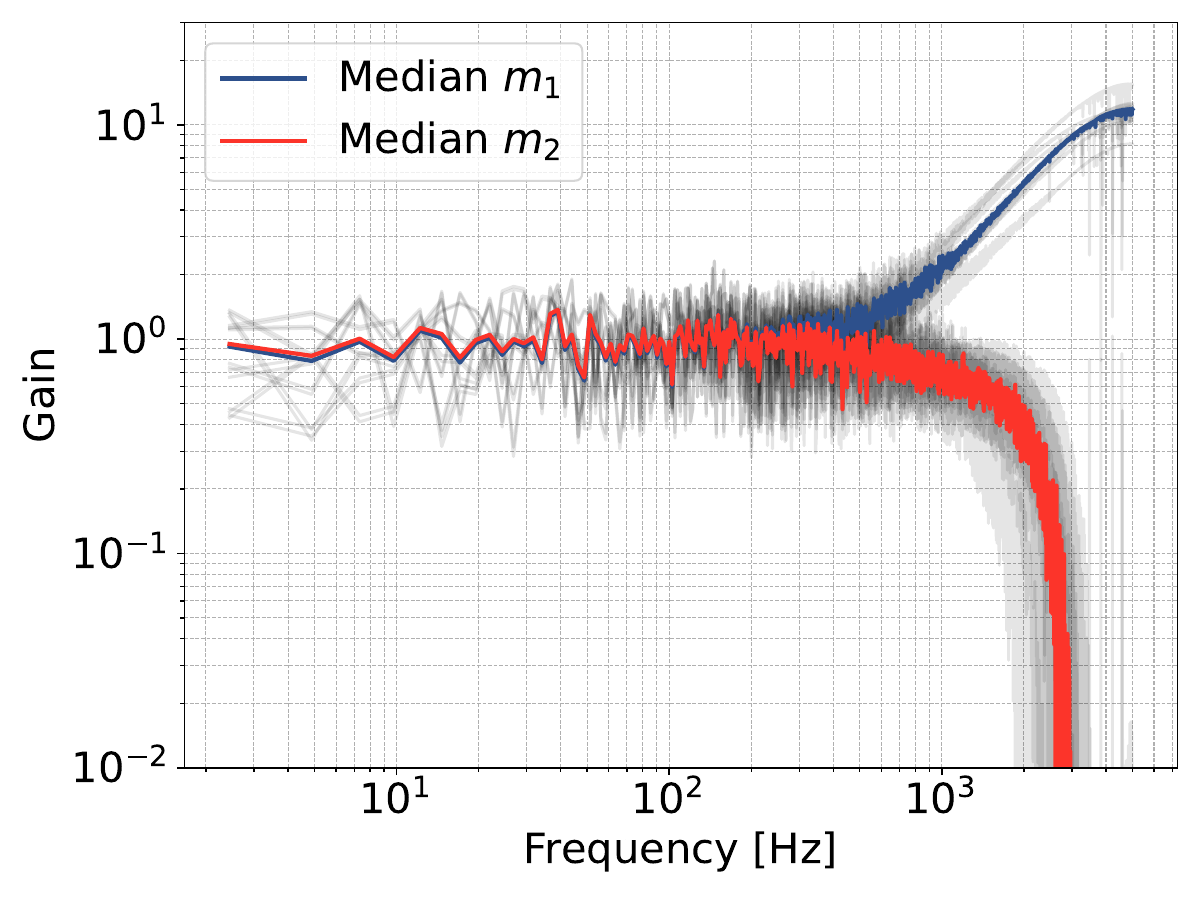}
    \caption{Optimized filter shapes $m_1(\omega)$ and $m_2(\omega)$ for the studied detectors at 90\% efficiency, along with the median shape computed across all detectors.}
    \label{fig:filters}
\end{figure}

The optimized filter shapes $m_1(\omega)$ and $m_2(\omega)$ are shown in Fig.~\ref{fig:filters}. 
The complementary frequency weighting reflects the discriminant design to maximize pile-up rejection while preserving signal efficiency.
Below 250~Hz, both functions have gains close to unity with less than 5\% variation across detectors.
$m_1(\omega)$ shows a first-order rising response ($\sim$22 to 29~dB/decade) in the 1--3~kHz range, amplifying the high-frequency features essential for resolving pile-up events, with maximum gains ranging from 8.2 to 15.5 across detectors.
In contrast, $m_2(\omega)$ reaches its minimum between 2.5~kHz and 3~kHz, above which gains fall below $10^{-3}$, providing complementary weighting that emphasizes the bulk signal shape over fast transients.
Different detectors show minor variations in their optimized filters due to individual signal and noise profiles, yet the core behavior remains consistent with individual filters exhibiting excellent agreement to the median shape (mean coefficient of determination $R^2 = 0.99$ for $m_1$ and $R^2 = 0.82$ for $m_2$).

The performance achieved for each detector is summarized in Table~\ref{tab:performance}, which reports the rise time, SNR, and BI obtained with the different pile-up rejection methods considered in this work. The analysis for this work was performed using a processing window of 0.4~s and grid step of 8~$\mu $s for the time separation $\Delta t$ and $5\times10^{-3}$ for the amplitude ratio $r$.
When comparing the pile-up BI obtained with the optimum filter amplitude ratio-based algorithm presented in \cite{CUPID_pileup} (ref.~method) to the method introduced here, we observe an average improvement of about 31\% for the same injected pulse shape (BI(i)). 
In a more-in depth systematics study (see Sec.~\ref{Sec:Systematics}) we found a significant systematic uncertainty associated with this pulse template used for the pile-up evaluation in \cite{CROSS_NTL}. As a more conservative choice we hence suggest to use an average pulse based BI prediction (BI (ii)) for the most probable absolute background index extrapolation going forward.

We note, that the reference pile-up rejection method \cite{CUPID_pileup} was also used in \cite{CROSS_NTL}, but that the resulting improvement quoted here differs compared to the published result in this article. When compared to the results in \cite{CROSS_NTL} the improvement would be $\sim21$\%.  
However, a more detailed investigation of these prior results revealed that the optimum filter discriminant in \cite{CROSS_NTL} contained a residual signal component in the noise power spectrum. This resulted in an effective transfer function that partially enhanced pile-up rejection at the expense of degraded amplitude estimation. To provide both well-defined optimum filter SNR and BI values, we recomputed BI and SNR with a clean noise power spectrum and provide those as (ref. method) in Table \ref{tab:performance}.

\begin{table}[h]
\centering
\setlength{\tabcolsep}{5pt}
\caption{Summary of detector performance for the pile-up discrimination method compared to the standard optimum filter at 90\% single-pulse efficiency. 
The values labeled BI (ref. method) correspond to the background indices obtained using the method described in~\cite{CUPID_pileup}, while those labeled BI~\cite{CROSS_NTL} are taken directly from ~\cite{CROSS_NTL}. Columns (i) and (ii) correspond to the BI from the analysis presented here with injected pulse templates defined in Sec.~\ref{Sec:Systematics}.
$E_\mathrm{temp}$ corresponds to the energy of the template pulse in the LD used for injection (the template is normalized prior to injection).
The BI is expressed in units of $10^{-5}$~cts/keV/kg/yr. 
Quoted uncertainties are only systematic, statistical uncertainty is $\pm 0.1$ for all detectors.}
\label{tab:performance}
\begin{tabular}{c c c c c c c c}
\hline
\textbf{Det.} & \textbf{Rise} & \textbf{SNR}\footnotemark & \textbf{$E_\mathrm{temp}$} &
\textbf{BI} &
\textbf{BI} & \textbf{BI} & \textbf{BI}\\
     & (ms) &     & (keV) & \makecell{(ref.\\method)} & \cite{CROSS_NTL} & (i) & (ii) \\
\hline
2  & 0.72 & 156 & 39 & 10.1 & 8.3 & $7.1^{\scriptscriptstyle +0.2}_{\scriptscriptstyle -0.4}$ & $7.3${\raisebox{0.5ex}{\tiny$^{\scriptscriptstyle+0.1}_{\scriptscriptstyle-0.6}$}}\\
3  & 0.54 & 115  & 96 & 10.0 & 8.3 & $6.8^{\scriptscriptstyle+0.2}_{\scriptscriptstyle-0.3}$ & $7.0^{\scriptscriptstyle+0.1}_{\scriptscriptstyle-0.4}$\\
4  & 0.55 & 246 & 44 & 6.6  & 5.9 & $4.5^{\scriptscriptstyle+0.4}_{\scriptscriptstyle-0.1}$ & $5.0^{\scriptscriptstyle+0.1}_{\scriptscriptstyle-0.4}$\\
5  & 0.78 & 292 & 32 & 6.7  & 6.6 & $4.7^{\scriptscriptstyle+0.4}_{\scriptscriptstyle-0.1}$ & $5.0^{\scriptscriptstyle+0.2}_{\scriptscriptstyle-0.3}$\\
9  & 0.64 & 248 & 39 & 7.7  & 7.3 & $5.6^{\scriptscriptstyle+0.2}_{\scriptscriptstyle-0.2}$ & $5.9^{\scriptscriptstyle+0.1}_{\scriptscriptstyle-0.4}$\\
10 & 0.51 & 145 & 111 & 7.6  & 6.3 & $4.7^{\scriptscriptstyle+0.5}_{\scriptscriptstyle-0.1}$ & $5.4^{\scriptscriptstyle+0.1}_{\scriptscriptstyle-0.5}$\\
11 & 0.51 & 140 & 80 & 8.2  & 6.9 & $5.5^{\scriptscriptstyle+0.4}_{\scriptscriptstyle-0.1}$ & $5.8^{\scriptscriptstyle+0.1}_{\scriptscriptstyle-0.4}$\\
12 & 0.47 & 134  & 134 & 7.7  & 6.9 & $5.5^{\scriptscriptstyle+0.2}_{\scriptscriptstyle-0.2}$ & $5.6^{\scriptscriptstyle+0.1}_{\scriptscriptstyle-0.3}$\\
\hline
\textbf{Avg.} & \textbf{0.59} & \textbf{184} & \textbf{72} & \textbf{8.0} & \textbf{7.0} & $\textbf{5.5}^{\scriptscriptstyle+0.3}_{\scriptscriptstyle-0.2}$  & $\textbf{5.9}^{\scriptscriptstyle+0.1}_{\scriptscriptstyle-0.4}$ \\
\hline
\end{tabular}
\end{table}
\footnotetext{The SNR has been recomputed with with an optimum filter with a clean noise power spectrum without the residual signal component present in \cite{CROSS_NTL}.}

\subsection{Systematics and training validation}
\label{Sec:Systematics}
\begin{figure}[!h]
    \centering
    \includegraphics[width=0.8\linewidth]{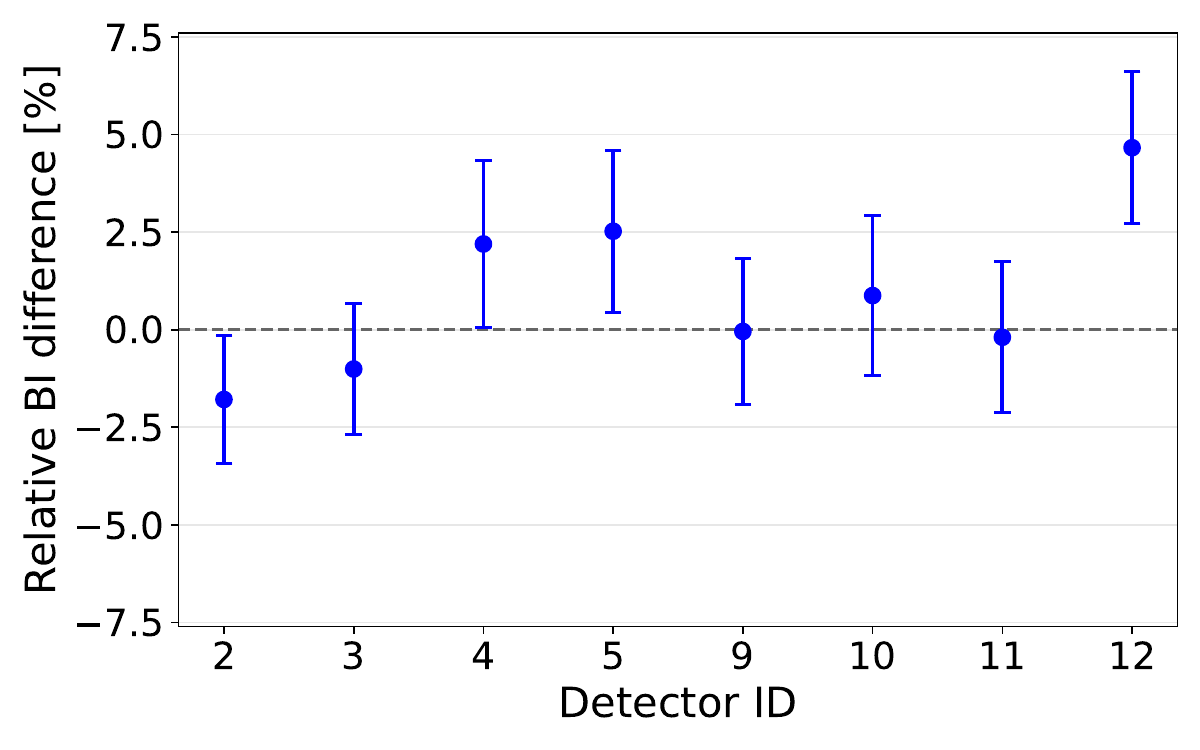}
    \caption{Comparison of BI obtained from simulated pile-up events versus the analytical model prediction.}
    \label{fig:simulated_vs_analytical_BI}
\end{figure}

To validate the analytical model (Sec.~\ref{sec:framework}), the BI derived from simulated pile-up events was compared with the predictions of the analytical framework. The comparison shows good agreement, with a detector-dependent bias that remains below 5\%. We note that we would expect minor violations of the assumptions of the analytical model like the assumption of perfectly uncorrelated time invariant noise to result in a slightly worse pile-up rejection performance of the simulated pulses with actual noise data in the evaluation data set. 
This seems to be confirmed in the data , where the results across detectors are presented in Fig.~\ref{fig:simulated_vs_analytical_BI}, with an average change in BI of $0.9\%$, albeit with considerable detector-to-detector variation. Nevertheless, the overall differences remain small confirming the reliability of the analytical approach as a numerically convenient way for filter training. To avoid any mistake caused by a violation of the assumptions used for the analytical computation (especially the noise having some non-stationary components), we opt to only use results from the pile-up pulses injected on real noise when reporting filter performance in terms of  experimental BI. In consequence we do not consider the observed difference between the analytical computation as a systematic on the BI listed in Table~\ref{tab:performance}.

\begin{figure}[!h]
    \centering
    \includegraphics[width=0.8\linewidth]{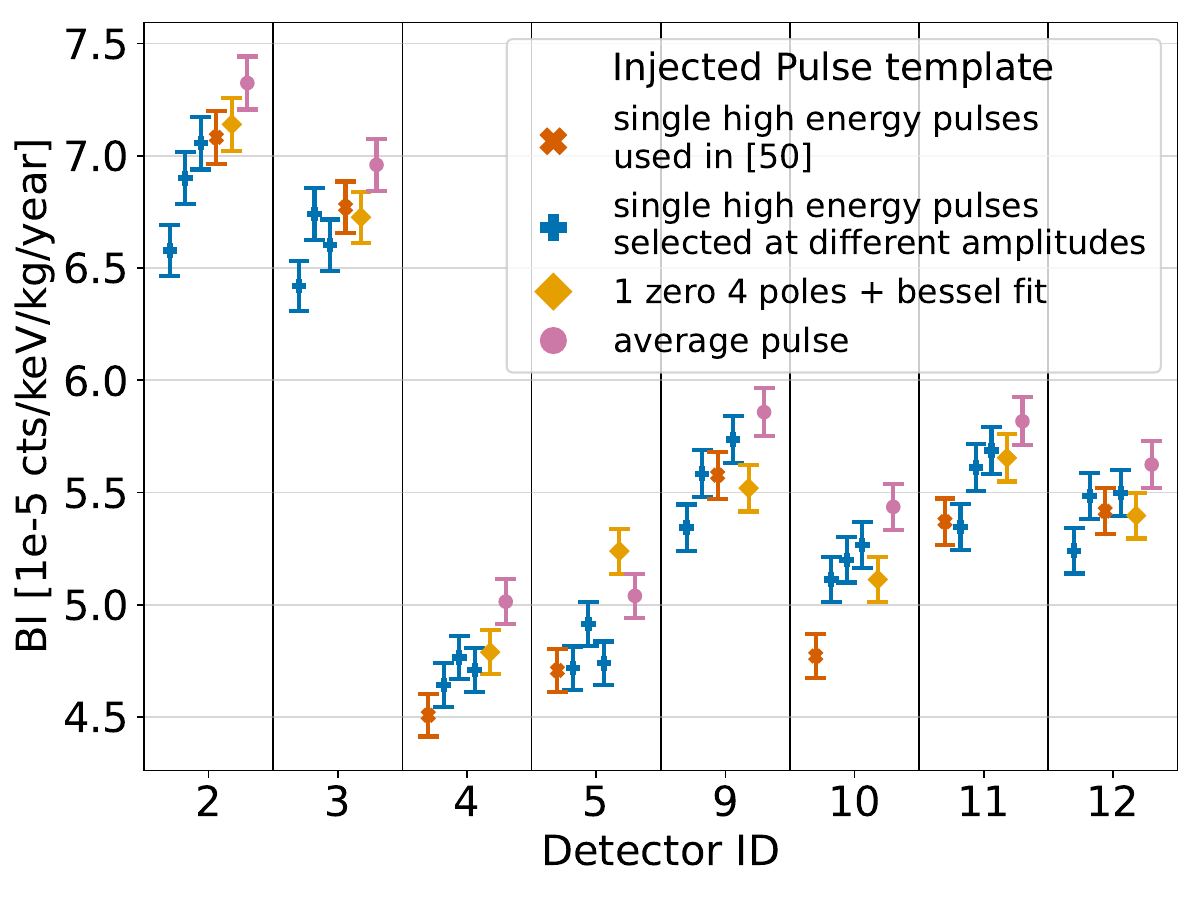}
    \caption{Comparison of different pulse templates used for pile-up discrimination on the BI: single high-energy pulses (sorted by amplitude), phenomenological model plus Bessel filter and average pulse.
}
    \label{fig:pulse_templates}
\end{figure}

The choice of the pulse template used to construct injected pile-up events can introduce systematic effects. Several template definitions were investigated, and the results are summarized in Fig.~\ref{fig:pulse_templates}. Three classes of templates were considered:  
(i) a single high-energy pulse, selected from different energy ranges (44~keV, 76~keV, 109~keV, and the template used in ~\cite{CROSS_NTL}, whose energy is reported in Table~\ref{tab:performance});  
(ii) an average pulse constructed from a large ensemble of events in the data, spanning a broad energy range from 4~keV to 306~keV with a median energy of 50~keV, due to the limited statistics available in narrower energy intervals;  
(iii) a phenomenological pulse model obtained by fitting a single high-energy pulse (the one used in ~\cite{CROSS_NTL}) with a function composed of one zero and four poles, convolved with a digital Bessel filter to reproduce the analog response of the acquisition chain~\cite{DAQ}.
Single-pulse templates generally yield the lowest BI, with a significant dependence on pulse energy such that higher-energy templates lead to larger background values, while the average-pulse template typically produces the highest BI. We believe that the different performance for single pulse templates is likely dominated by a residual high-frequency noise imprint in single pulse templates, which is an artefact that may be exploited by the optimized weighting functions that emphasize high-frequency contributions (see Fig.~\ref{fig:filters}). 
The overall variation in BI across the different templates reaches up to 13\%. The choice of template should therefore be carefully considered in future applications. The use of the average pulse template (ii) is recommended as a conservative option, as it preserves the characteristic pulse shape observed in data, while avoiding the injection of excess high-frequency noise present in single high-energy pulses. However, imperfections in the alignment may broaden the averaged pulse template and are hence expected to slightly deteriorate the estimated pile-up rejection performance. An energy dependence of the detector response with a template chosen far from the expected energy ($\sim$ 1 keV ) of a pulse is a further possible source of systematics to be studied in the future.

The choice of the time window used for pulse processing can influence the pile-up discrimination performance. By varying the window length between 0.1~s and 1~s, the resulting BI exhibits the expected dependence, with longer windows providing a modest improvement in SNR and discrimination power due to enhanced frequency resolution. Performance reaches a plateau for window lengths greater than 0.3~s, where the residual variation in the BI remains below 3.3\%. The position of the pulse within the processing window, using an asymmetric windowing resulted in a variation of the BI of up to 1.1\%.

\begin{figure}[!h]
    \centering
    \includegraphics[width=0.8\linewidth]{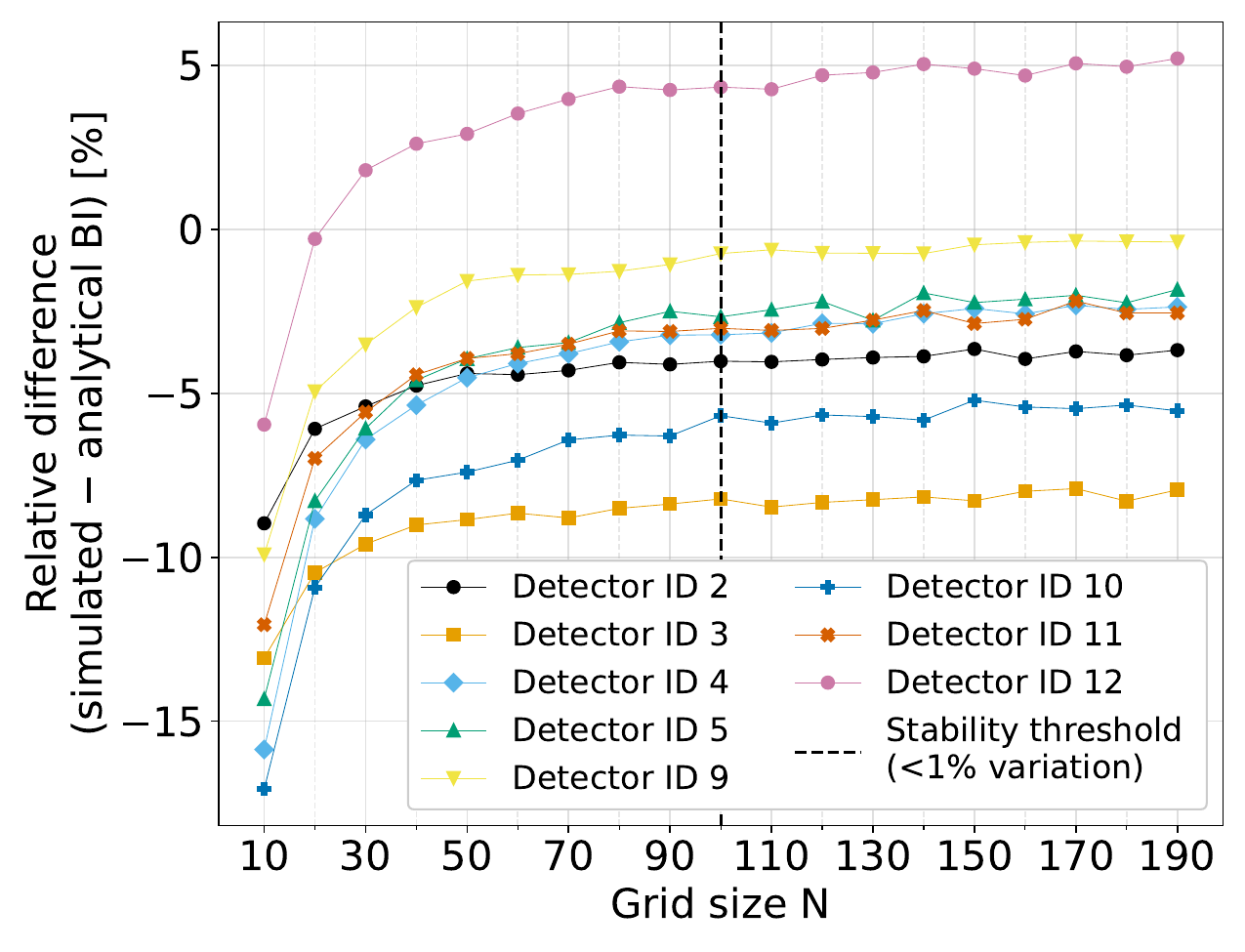}
    \caption{Impact of the time separation $\Delta t$ and amplitude ratio $r$ grid size on the BI obtained from simulated pile-up events and from the analytical model prediction. The grid size $N$ is defined as the number of steps in each dimension for $\Delta t \in [0, 8\times10^{-4}]$~s and $r \in [0, 0.5]$.}
    \label{fig:grid_size_systematic}
\end{figure}

We investigated the impact of the discretization of the time separation $\Delta t$ and amplitude ratio $r$ used in the cost function by varying the grid resolution. Over the explored range, shown in Fig.~\ref{fig:grid_size_systematic}, the trained filter functions $m_1$/$m_2$ remained stable with a BI variation of less than 0.6\% evaluated from simulated pile-up events, indicating a weak dependence on the grid choice. In contrast, the analytical BI prediction exhibited a stronger sensitivity to the grid resolution, reaching a stability level of about 1\% for $\Delta t$ steps smaller than 8~$\mu$s and $r$ steps below 0.005 (N = 100).
Since the computational complexity of the optimization scales as $N^2$, a reduced grid size can enable rapid feedback during detector working point tuning and optimization to provide a viable relative comparison. On the other hand the chosen grid size of 100 steps, ensures that the BI estimation remains within 1\% of its asymptotic value.
For the systematics quoted in Table~\ref{tab:performance}, the four different effects summarized in Table~\ref{tab:systematics_summary} (Pulse template, processing window, pulse position and training grid size) are treated as independent and added in quadrature.

\begin{table}[!h]
\centering
\caption{Summary of the systematics for the pile-up BI estimation.}
\label{tab:systematics_summary}
\begin{tabular}{l c}
\hline
\hline
Source & BI variation \\
\hline
Pulse template (injection) & up to $13\%$ \\
Processing window size ($>0.3$\,s) & $<3.3\%$ \\
Pulse position in window & $<1.1\%$ \\
Grid discretization ($\Delta t$, $r$) & $<0.7\%$ \\
\hline
\hline
\end{tabular}
\end{table}
\subsection{Detector operation point dependence}

The detectors are read out using Neutron Transmutation Doped (NTD) germanium thermistors \cite{NTD}, whose resistance is measured by biasing them with a constant current. At fixed base temperature, the bias current determines the equilibrium temperatures of the electron and phonon systems of the sensor, and therefore affects both the sensitivity and the signal response, as well as the pulse shape through temperature-dependent heat capacities and thermal conductance. Variations of the dynamic impedance can also introduce secondary effects due to RC filtering along the readout lines.

A detailed study of the full electro-thermal response of the detectors, as performed for \teo calorimeters in~\cite{CUORE_thermal_response}, is beyond the scope of this work. Here, instead we present a pragmatic first test of the dependence of pile-up background rejection on the detector working point, using a very limited dataset taken in \cite{CROSS_NTL}  and the proposed framework.

The data taking in~\cite{CROSS_NTL} was primarily devoted to the study of the overall performance of NTL-assisted LDs at the tower level and only limited data were acquired for the optimization of individual detector working points. In particular, only two devices were operated in a scan over different bias currents ranging between 0.3 and 10~nA (corresponding to resistances between 0.6 and 26.2~M$\Omega$) to determine their SNR and rise time as proxies for pile-up performance.

\begin{figure}[!ht]
    \centering
    \includegraphics[width=0.8\linewidth]{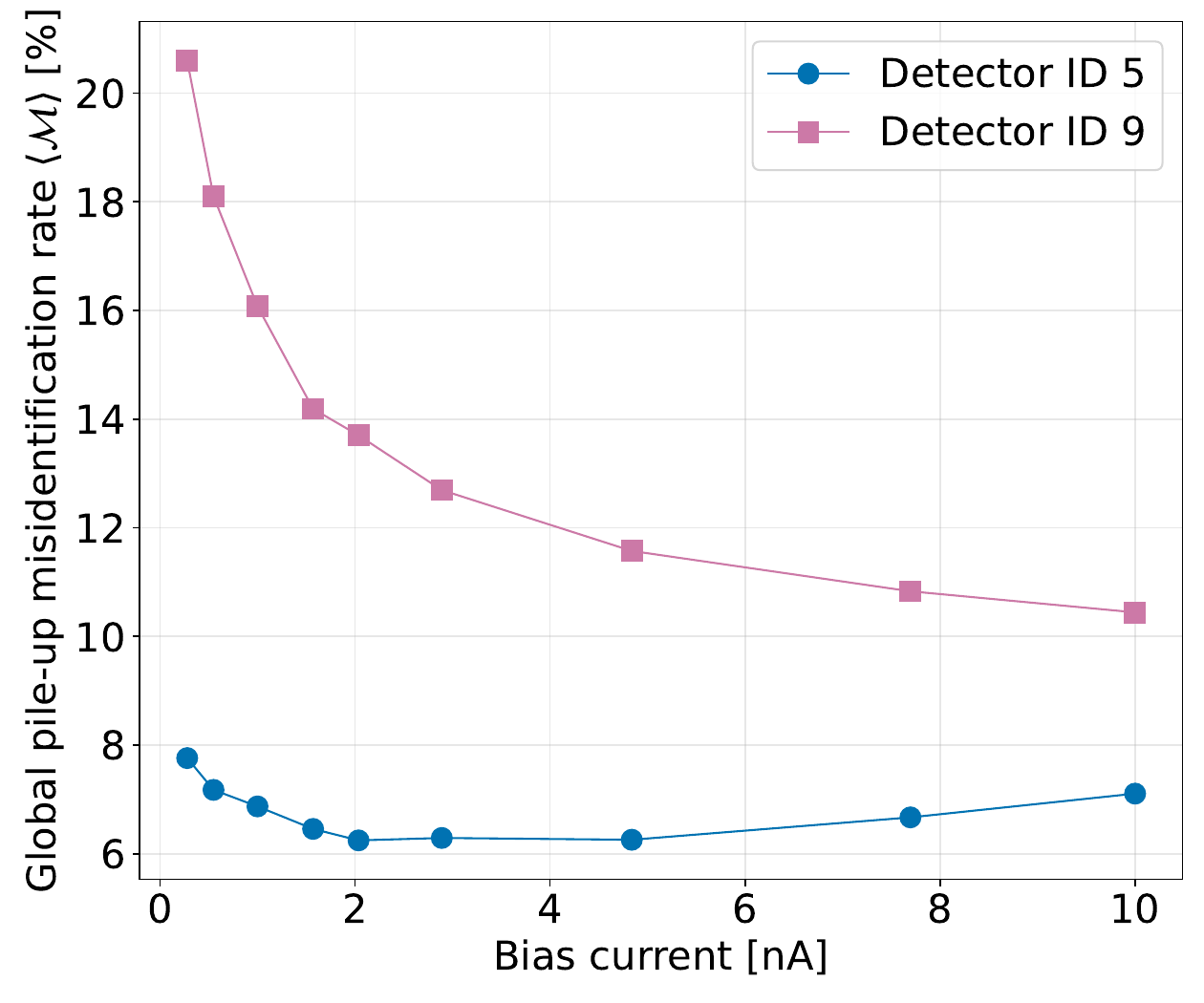}
    \caption{Impact of different working points of the NTD Ge thermistor (constant current bias) on the global pile-up misidentification rate.}
    \label{fig:WP_test}
\end{figure}

Although these data were insufficient  (short measurements) to perform a full pile-up rejection analysis based on the injection of synthetic pulses into individual noise traces, they allowed the extraction of average noise power spectra and average pulse templates for each working point. This enabled the optimization of the pile-up filter and the prediction of the pile-up misidentification rate for each bias current using the algorithm presented above.
Figure~\ref{fig:WP_test} shows the resulting dependence of the global pile-up misidentification rate on the bias current for the two detectors included in the scan. For detector~5, a minimum is observed in the range between 2 and 5~nA, with an overall variation in background rejection of about 14\% across the explored bias range. In contrast, detector~9 exhibits a much stronger dependence, with variations reaching a factor of two and no minimum observed within the investigated range.
Despite being limited by both the short measurement time and the small number of detectors, these results indicate that a non-negligible improvement in pile-up rejection may be achieved through an individual optimization of the detector working point similar to the individual working point optimization already performed for the \lmo detectors. In this context, the algorithm presented in this work represents a valuable tool: it is computationally efficient and requires only limited data to construct a signal template and noise power spectrum for each working point. This makes it well suited to fully account for changes in SNR, pulse shape, and noise properties when selecting the detector operating conditions.

\section{Discussion \& Outlook}
\label{sec:outlook}

In this work, we have introduced a general framework for pile-up discrimination based on the construction of an optimized filter in the frequency domain. The formalism assumes stationary pulse shapes, stationary noise fully described by its power spectral density, and a fixed target efficiency for single-pulse events. Under these assumptions, the optimization procedure determines the filter that minimizes the pile-up misidentification rate for the chosen discriminant parameter.

A central feature of the present approach is the use of a ratio-based discriminant built from two filtered observables. This choice was motivated by robustness, interpretability, and its weak dependence on absolute signal amplitude. The optimization exploits the strong correlation between numerator and denominator to reduce sensitivity to amplitude fluctuations and improve discrimination power. 
However, the discrimination remains confined to a one-dimensional projection of the waveform information. Alternative parameterizations, such as differences or higher-order combinations of filtered observables, may extract additional information and further enhance performance.
An additional, though less fundamental, limitation arises from constraining the frequency-domain weighting functions to be real and positive. 
This choice ensures the stability of the optimization and preserves a clear physical interpretation of the weighting function as a frequency-dependent gain, while avoiding pathological solutions driven by noise-dom\-inated frequency regions. Nevertheless, it restricts the space of admissible filters. Allowing complex-valued or sign-chang\-ing weighting functions could in principle exploit phase information or interference effects between frequency components and might therefore improve discrimination in more general settings. Such extensions would require a careful reassessment of the statistical interpretation of the discriminant and of its robustness against noise fluctuations.

Beyond these intrinsic limitations, the method relies on several assumptions that define its practical domain of validity. The noise is assumed to be stationary and Gaussian, and the pulse shape is treated as stable and representative of all events in the region of interest. Deviations from these assumptions, such as pulse shape variations with energy or time, non-Gaussian noise components, or slow drifts in detector response, introduce systematic effects that can bias the optimization.
Despite these limitations, the method demonstrates a significant and robust reduction of the pile-up background across a set of detectors, with a median improvement of approximately 31\% at 90\% signal efficiency using the same pulse template as in \cite{CROSS_NTL} and of 26\% using a pulse template resulting in more conservative results. 
This indicates that the ratio-based discriminant captures the dominant pile-up features relevant for cryogenic calorimeters with slow thermal response and enables an efficient discrimination of pile-up events primarily induced by two-neutrino double-beta decays of $^{100}$Mo. 
The extrapolated mean residual background index of $5.9 ^{\scriptscriptstyle+0.1}_{\scriptscriptstyle-0.4} \times 10^{-5}$~cts/keV/kg/yr (median: $5.7 \times 10^{-5}$ cts/keV/kg/yr) approaches the CUPID design goal of a pile-up contribution below $5.0 \times 10^{-5}$ cts/keV/kg/yr. 
Ongoing hardware developments aimed at the full coverage of the NTL amplification region (configuration IV in \cite{CROSS_NTL}) would result in a  background index of $5.3 \times 10^{-5}$~cts/keV/kg/yr (median: $5.1 \times 10^{-5}$ cts/keV/kg/yr) with the novel analysis technique presented here. Using this technique, the measured performance of the square CROSS light detectors are expected to result in a background of $7.7 \times 10^{-5}$~cts/keV/kg/yr (median: $7.5 \times 10^{-5}$ cts/keV/kg/yr)  in CROSS (\cite{CROSS_NTL} configuration I), sub-dominant compared to other backgrounds in the demonstrator \cite{CROSS_SENSITIVITY}.
We note though that the extrapolation for CUPID is particularly sensitive to the noise environment in the O(1) kHz range and that CROSS and CUPID have different mechanical designs and isolation mechanism for vibrational noise sources. 
CUPID's cryogenic infrastructure will undergo dedicated upgrades in 2027 to improve, assess and provide in-situ monitoring of its noise environment. 
A more complete extrapolation of the expected pile-up background after the upgrade will need to take these improvements into account. 
In parallel ongoing hardware developments, like an optimization of the light detector electrode spacing and geometry, and quality assurance/pre-testing protocols aim at improving the basic performance of Ge or Si based LD's for CUPID in terms of rise-times and in particular SNR. If successful these will result in further pile-up rejection improvements.

Future developments on the analysis methodology could explore alternative discriminant constructions, including multi-parameter or multivariate approaches, while retaining the analytically grounded optimization strategy presented here. As one example a future analysis could make use of the overall information of both the \lmo readout channel and its two LD's (above and below). 
Alternatively, allowing for complex weighting functions may allow to more fully exploit phase differences and improve pile-up rejection with the present discriminant on a single detector basis.
Finally, extending beyond the amplitude-ratio based discriminant to systematically compare different classes of discriminants on equal footing and leveraging machine learning techniques would allow a more global assessment of optimality and may further enhance pile-up rejection in next-generation cryogenic experiments.

\begin{acknowledgements}
This work is supported by the European Commission (Project CROSS, Grant No. ERC-2016-ADG, ID 742345) and by the Agence Nationale de la Recherche (ANR France; Project CUPID-1, ANR-21-CE31-0014).
\end{acknowledgements}

\bibliographystyle{spphys}       
\bibliography{Bibliography.bib}   

\appendix

\section{Computation of Statistical Moments used in the Discriminant Parameter}
\label{app:stat_moment}

This appendix provides the derivations of the statistical properties of the pile-up discriminant parameter Y~($\Delta$t, r) within the framework and assumptions of the pile-up model, as introduced in Sec.~\ref{sec:framework}.

Assuming that the noise is small and has zero-mean, the expected values of $X[m_i]$ can be approximated as:
\begin{equation}
    \mu_i(\Delta t,r) \approx A \max \operatorname{Re} \Big[ \mathcal{F}^{-1} \big( m_i(\omega) H(\omega) \hat{s}(\omega)(1-r+ r e^{i\omega \Delta t}) \big) \Big].
\end{equation}

The covariance between \(X[m_1]\) and \(X[m_2]\) is defined as
\begin{equation}
\operatorname{Cov}(X[m_1], X[m_2]) = \mathbb{E} \big[ (X[m_1] - \mu_1)(X[m_2] - \mu_2) \big].
\end{equation}

Using the definition of the real part, \(\operatorname{Re}(z) = \frac{1}{2}(z + z^*)\), we can write
\begin{equation}
\begin{aligned}
\operatorname{Cov}(X[m_1], X[m_2])
&= \frac{1}{4}\,\mathbb{E}\Bigg[ \frac{1}{2\pi}
\biggl( \int m_1(\omega_1)
\Bigl(
  H(\omega_1)\hat n(\omega_1) \\
&\qquad\qquad
+ H^*(\omega_1)\hat n^*(\omega_1)
\Bigr)
\, d\omega_1 \biggr) \\
&\qquad\qquad \times  \frac{1}{2\pi}
\biggl( \int m_2(\omega_2)
\Bigl(
  H(\omega_2)\hat n(\omega_2) \\
&\qquad\qquad
+ H^*(\omega_2)\hat n^*(\omega_2)
\Bigr)
\, d\omega_2 \biggr)
\Bigg].
\end{aligned}
\end{equation}

Since \(\hat{n}(\omega)\) is zero-mean, only second-order moments contribute, so considering two separated frequencies $\omega$ and $\omega'$:
\[
\mathbb{E}[\hat{n}(\omega) \hat{n}(\omega')] = 0, \quad
\mathbb{E}[\hat{n}^*(\omega) \hat{n}^*(\omega')] = 0, \quad
\]
\[
\mathbb{E}[\hat{n}(\omega) \hat{n}^*(\omega')] = \mathcal{P}_N(\omega) \, \delta(\omega - \omega').
\]

Plugging these relations in, the covariance simplifies to:
\begin{equation}
\begin{split}
\operatorname{Cov}(X[m_1], X[m_2]) 
&= \frac{1}{4}  \Big(\frac{1}{2\pi}\Big)^2 \int m_1(\omega) m_2(\omega) \Big( 
      H(\omega) H^*(\omega) \\
&\quad + H^*(\omega) H(\omega) \Big) \mathcal{P}_N(\omega) \, d\omega \\
&= \frac{1}{2}  \Big(\frac{1}{2\pi}\Big)^2 \int m_1(\omega) m_2(\omega) |H(\omega)|^2 \\ 
&\quad \mathcal{P}_N(\omega) \, d\omega.
\end{split}
\end{equation}

Taking $m_1=m_2$, we get that the variances are
\begin{equation}
\begin{split}
\sigma_i^2(\Delta t,r) = \frac{1}{2} \Big(\frac{1}{2\pi}\Big)^2  \int m_i^2(\omega) |H(\omega)|^2 \mathcal{P}_N(\omega) \, d\omega, \\
\quad i = 1,2.
\end{split}
\end{equation}

Using a small-variance approximation for the ratio of correlated Gaussians \cite{Normal_ratio}, the mean and variance of $Y(\Delta t,r)$ are:
\begin{align}
    \mu_Y[m_1,m_2](\Delta t,r) &= \frac{\mu_1(\Delta t,r)}{\mu_2(\Delta t,r)},\\
    \sigma_Y^2[m_1,m_2](\Delta t,r) &\simeq \frac{\sigma_1^2}{\mu_2^2} + \frac{\mu_1^2}{\mu_2^4} \sigma_2^2 - 2 \frac{\mu_1}{\mu_2^3} \operatorname{Cov}(X[m_1],X[m_2]).
\end{align}


\end{document}